\documentclass[pra,lengthcheck,nofootinbib,amsmath,amssymb,aps,floatfix]{revtex4-2}
\usepackage[utf8]{inputenc}
\usepackage{graphicx}
\usepackage[dvipsnames]{xcolor}
\usepackage[caption=false]{subfig}
\usepackage{bm}
\usepackage{physics}
\usepackage[colorlinks,allcolors=blue]{hyperref}

\def\equationautorefname#1#2\null{Eq.#1(#2\null)}

\newcommand{\appref}[1]{\hyperref[#1]{\appendixautorefname~\ref*{#1}}}

\renewcommand{\Im}{\operatorname{Im}}

\renewcommand{\vec}[1]{\boldsymbol{\mathbf{#1}}}

\allowdisplaybreaks{}

\begin{document}

\title{High-harmonic generation driven by temporal-mode quantum states of light}

\newcommand{\FTMC}{Departamento de Física Teórica de la Materia Condensada, Universidad Autónoma de Madrid, E-28049 Madrid, Spain}
\newcommand{\IFIMAC}{Condensed Matter Physics Center (IFIMAC), Universidad Autónoma de Madrid, E-28049 Madrid, Spain}
\newcommand{\INC}{Instituto Nicolás Cabrera (INC), Universidad Autónoma de Madrid, E-28049 Madrid, Spain}

\author{Juan M. González-Monge}
    \email{juanmanuel.gonzalezm@estudiante.uam.es}
    \affiliation{\FTMC}
    \affiliation{\IFIMAC}
    \affiliation{\INC}
\author{Felipe Reibnitz Willemann}
    \affiliation{\FTMC}
    \affiliation{\IFIMAC}
    \affiliation{\INC}
\author{Johannes Feist}
    \email{johannes.feist@uam.es}
    \affiliation{\FTMC}
    \affiliation{\IFIMAC}
    \affiliation{\INC}
\date{\today}

\begin{abstract}
We develop a theoretical framework for high-harmonic generation (HHG) driven by quantum states of light based on a temporal-mode expansion of the electromagnetic field. This approach extends previous single plane-wave mode treatments to realistic pulse configurations and arbitrary multi-mode states of light, resolving conceptual inconsistencies arising from non-normalizable infinite plane waves and establishing consistency between analytical and numerical methods. We derive a correction factor that quantifies deviations from the diagonal approximation (in which the yield becomes a statistical average over classical-field simulations) both for the response of a single atom and in the many-atom regime. Our results confirms that the HHG spectrum for atoms driven by any quantum state of light in free space is accurately described by averaging semi-classical calculations over the Husimi distribution, with no observable genuine quantum effects in the spectrum. We also demonstrate that in the many-atom regime, the mean-field coherent-state approximation underlying this treatment does not preserve probabilities, although unitarity is restored by in the diagonal approximation.
The absence of genuine quantum effects in the HHG yield is attributed to the large photon numbers ($\sim 10^{11}$) required to reach HHG intensities in free space, which render quantum fluctuations negligible. We discuss nanophotonic environments with ultrasmall mode volumes as potential platforms where few-photon strong-field processes could exhibit genuine quantum signatures.
\end{abstract}

\maketitle

\section{Introduction}
\label{sec.introduction}
In the last few years, there has been increasing interest in the quantum features of light in the context of attosecond science and strong-field physics~\cite{Gonoskov2016,gorlach2020,Gombkoto2021,Tzur2023,Stammer2023,Gonoskov2024,Yi2025,Tsatrafyllis2017,lewenstein2021,Theidel2024,Theidel2025}. 
An important motivation for this has been the development of bright squeezed vacuum (BSV) sources~\cite{Iskhakov2009,Spasibko2012,Chekhova2015}, which enable the use of nonclassical driving pulses for strong-field physics~\cite{gorlach2023,Rasputnyi2024,Liu2025Atomic,Mouloudakis2018,Mouloudakis2019,Lemieux2025,Tzur2025}.
Studying the process of high-harmonic generation (HHG), Gorlach et al.\ developed a theoretical framework demonstrating that, under reasonable approximations, the HHG signal driven by quantum states of light can be obtained by averaging several \emph{semi-classical} calculations with classical driving fields~\cite{gorlach2023}.
The weights in the averaging are determined by the Husimi quasi-probability distribution $Q(\alpha) = \expval{\rho_\mathrm{F}}{\alpha}$, where $\rho_\mathrm{F}$ is the density operator of the electromagnetic field and $\ket{\alpha}$ is a coherent state, with the classical field amplitude directly proportional to the coherent state amplitude $\alpha$.

This prescription implies that HHG driven by any quantum field can be equivalently reproduced by a statistical average over a collection of simulations with classical pulses, indicating that there can be no ``quantum advantage'' in HHG\@.
The increased cutoff observed for quantum states of light such as BSV is simply due to the fact that the electric field amplitude fluctuates strongly in these states. Consequently, for a given average intensity, much higher peak intensities are included in the statistical distribution.
A collection of classical pulses with different peak intensities (and the same intensity distribution) would provide the same enhancement in the cutoff.

However, the derivation in Ref.~\cite{gorlach2023} relies on several assumptions that may not hold in all situations.
In particular, the driving field is assumed to be in a single-mode plane-wave state even in the limit of infinite volume, which means that it fills all of space and is, in principle, not in a normalizable state.
The derivation depends on the fact that the formal normalization prefactor $\propto V^{-1/2}$ vanishes in the free-space limit $V \to \infty$, where $V$ is the quantization volume.
Furthermore, the numerical calculations do not actually employ an infinite plane wave, but rather a finite one with a well-defined temporal profile with a total duration of 25 laser cycles.
The numerical and analytical results are thus not fully consistent with each other.

In the current work, we resolve these issues by extending the framework developed in Ref.~\cite{gorlach2023} to the case of a driving field that is not in a single plane-wave state, but in a temporal mode—i.e., a wavepacket that is localized in time and space.
Generalizing the results obtained in Ref.~\cite{gorlach2023} to this more realistic scenario, we obtain an analytical result almost identical to the one given there, with a correction factor that can be shown to be extremely close to unity for realistic driving fields containing many photons.

Our results also close a small loophole in the argument given above: Since the ``most classical'' state of light, a coherent state, is itself a quantum state with non-zero width in the Husimi distribution (i.e., with unavoidable quantum noise), even HHG driven by a coherent field must include some averaging over different classical field amplitudes.
Consequently, non-classical states of light (whose Husimi functions can have features narrower than a unit width) could still potentially exhibit genuine quantum effects in the associated HHG spectrum.
Our results demonstrate explicitly that these effects are negligible for all realistic driving fields in free space, precluding the observation of genuine quantum effects due to the statistics of the driving pulse, unless extremely strong subwavelength confinement of the driving field is achieved such that the intensities required for HHG are reached with a small number of photons.

The manuscript is structured as follows: We first introduce the theoretical framework in \autoref{sec.theory}, where we derive the main result of this work. In \autoref{sec.results}, we present numerical results for HHG driven by different quantum states of light and compare the results obtained using the single-mode and temporal-mode frameworks. Finally, we summarize our findings and their implications in \autoref{sec.Summary}.

\section{Theoretical framework}
\label{sec.theory}

We treat a single-electron atom interacting with the quantized electric field within the framework of non-relativistic quantum electrodynamics, employing the dipole (long-wavelength) approximation in the Power-Zienau-Woolley picture. The Hamiltonian describing the system consists of three terms~\cite{Feist2021}:
\begin{equation}
    H = H_\mathrm{A} + H_\mathrm{F} + \vec{d}_\mathrm{A} \cdot \vec{E}(\vec{r}_\mathrm{0})
    \label{eq:Hamiltonian}
\end{equation}
where $H_A = \frac{\Vec{p}^2 }{2m} + U(\Vec{r})$ is the atomic Hamiltonian describing an electron with momentum $\vec{p}$ moving in the potential $U(\vec{r})$ of the nucleus (assumed fixed in space at $\vec{r}=\vec{r}_0 = \vec{0}$), $H_F = \sum_{\ell} \hbar \omega_\ell a_{\ell}^{\dagger} a_{\ell}$ is the free-field Hamiltonian, and $\vec{d}_\mathrm{A} = e\vec{r}$ is the atomic dipole operator. The electric field operator $\vec{E}(\vec{r})$ expanded in a plane-wave basis is given by
\begin{subequations}
\label{eq:E_planewaves}
\begin{align}
    \Vec{E}(\Vec{r}) &= \sum\limits_{\ell} \vec{f}_\ell(\vec{r}) a_\ell + \mathrm{H.c.}\\
    \vec{f}_\ell(\vec{r}) &= i \Vec{e}_\ell \sqrt{\frac{\hbar \omega_\ell}{2 V \epsilon_0}} e^{i\vec{k}_\ell \cdot \vec{r}},
\end{align}
\end{subequations}
where $\ell = (\Vec{k}_\ell, \sigma_\ell)$ is a combined index running over both the wave vector $\Vec{k}_\ell$ and the polarization $\sigma_\ell$ (with two possible polarization states per wave vector) for compact notation, $\Vec{e}_\ell$ is the polarization vector, ${a}_{\ell}$ and ${a}_{\ell}^{\dagger}$ are the annihilation and creation operators, respectively, and $V$ is the quantization volume.
The frequency $\omega_\ell$ for each mode is given by $\omega_\ell = c|\vec{k}_\ell|$.

We write a sum over the mode index for simplicity of notation—in the continuous ($V\to\infty$) limit, this sum will become a mixture of integrals and sums.
Furthermore, in principle the atomic Hamiltonian should contain a polarization self-energy term that formally diverges but partially cancels the equally divergent contribution of the vacuum field to the light-matter interaction, producing a logarithmically divergent result that gives the (small) Lamb shift after proper regularization~\cite{Buhmann2012I}.
We do not explicitly include the polarization self-energy as we do not calculate the vacuum correction to the energy levels.

For any initial (mixed or pure) state, the von Neumann equation 
\begin{equation}
    i\hbar  \pdv{\rho(t)}{t} = [H, \rho(t)],
    \label{eq:Neumann}
\end{equation}
governs the time evolution of the density matrix $\rho(t)$ describing both the atom and the light field. For convenience, we switch to the interaction picture with respect to the field Hamiltonian $H_F$ via the transformation $\rho(t) \to e^{-\frac{i}{\hbar} H_\mathrm{F} t}\rho(t)e^{\frac{i}{\hbar} H_\mathrm{F} t}$. The equation to solve in the interaction picture is
\begin{equation}
    i\hbar  \pdv{\rho(t)}{t}  = [{H_o}(t), \rho(t)],
    \label{eq:density_equation}
\end{equation}
where ${H_o}(t) = H_\mathrm{A} + {\Vec{d}} \cdot {\Vec{E}}(\Vec{r},t)$, and the electric field now takes the form
\begin{subequations}
\label{eq:Interaction_field}
\begin{align}
    \Vec{E}(\Vec{r},t) &= e^{-\frac{i}{\hbar} H_\mathrm{F} t}{\Vec{E}}(\Vec{r}) e^{\frac{i}{\hbar} H_\mathrm{F} t} = \sum\limits_{\ell} \vec{f}_\ell(\vec{r},t) a_\ell + \mathrm{H.c.}\\
    \vec{f}_\ell(\vec{r},t) &= \vec{f}_\ell(\vec{r}) e^{-i\omega_\ell t} = i \Vec{e}_\ell \sqrt{\frac{\hbar \omega_\ell}{2 V \epsilon_0}} e^{i\left(\vec{k}_\ell \cdot \vec{r} - \omega_\ell t\right)}.
\end{align}
\end{subequations}

\subsection{Temporal mode basis}
We now perform a unitary transformation from the plane-wave basis to a basis of wavepackets or ``temporal modes''~\cite{Brecht2015}, $a_\ell = \sum_{\nu} U_{\ell\nu} b_{\nu}$, described by operators $b_{\nu}$. Therefore, the modes associated to the operators $b_{\nu}$ represent arbitrary linear combinations of the original plane-wave modes. Since the transformation is unitary, the resulting basis again describes independent bosons, i.e., $[a_\ell,a_{\ell'}^{\dagger}] = \delta_{\ell,\ell'}$ implies $[b_{\nu},b_{\nu'}^{\dagger}] = \delta_{\nu,\nu'}$. Inserting this transformation into the electric field operator yields
\begin{equation}
\begin{aligned}
   {\Vec{E}}(\vec{r},t) &= \sum\limits_{\nu} \vec{g}_{\nu}(\vec{r},t) {b}_{\nu} + \mathrm{H.c.},
\label{eq:E_introductionb}
\end{aligned}
\end{equation}
where we have defined the temporal mode functions $\vec{g}_{\nu}(\vec{r},t) = \sum_{\ell} U_{\ell\nu} \vec{f}_\ell(\vec{r},t)$. We note that the unitarity of $U_{\ell\nu}$ implies a normalization condition for the mode fields $\vec{g}_\nu(\vec{r},t)$.

In the following, we assume that the quantum driving field has nonzero contributions only within a finite subset of $k$ temporal modes. We denote the indices $\nu$ that span this space of driving modes by the composite index $\vec{\mathrm{p}} \equiv (\mathrm{p}_1, \ldots, \mathrm{p}_k)$.
This assumption replaces the single plane-wave state assumption in Ref.~\cite{gorlach2023} and decouples the state of the light field from the formal free-space limit $V \to \infty$.
It also resolves the conceptual problem of calculating the action of an infinitely extended (always-on) light field on a single atom, since we can choose the temporal modes $\vec{g}_\mathrm{p_i}(\vec{0},t)$ to have finite temporal support at the position of the atom, and make the numerical simulations consistent with the analytical results.
We note that any state of the EM field can of course be represented in any basis, and when treating arbitrary multi-mode driving fields, one could argue that just working in the original plane-wave basis would be conceptually simpler.
However, temporal modes are a practical way to separate the ``classical'' mode properties, described by the temporal mode functions $\vec{g}_\nu(\vec{r},t)$, from the ``quantum'' statistics, described by the quantum state of that mode, especially in the case where only a few temporal modes are sufficient to represent the field.
Indeed, experimental efforts directed at creating short-pulse quantum light pulses typically try to approach the single-temporal-mode limit~\cite{Law2000,Law2004,Rasputnyi2024,Demontigny2026}.
We therefore develop the theory within the framework of temporal modes without loss of generality.
Our assumption implies that in the initial state, only the modes of the set $\{b_\mathrm{p_i}\}_{i=1}^m$ will have non-zero photon population.
The associated mode functions $\{\vec{g}_{\mathrm{p_i}}(\vec{r},t)\}_{i=1}^m$ contain the information about the temporal shape and (by Fourier transform) frequency content of each involved mode, while the quantum statistics of the light is contained within the (potentially entangled) quantum state of these modes and probed by the operators $b_\mathrm{p_i}$.

\subsection{Time evolution}
We follow a procedure similar to that presented by Gorlach et al.~\cite{gorlach2023} to rewrite (under appropriate approximations) the general solution for the system density matrix in a form that is amenable to numerical solution.

Before the action of the light pulse, the electromagnetic field and atomic states are not entangled. Consequently, the initial density matrix $\rho(0)$ can be represented as the tensor product of the density matrices for the atom and the electromagnetic field: $\rho(0) = \rho_\mathrm{A}(0) \otimes \rho_\mathrm{F}(0)$, where $\rho_\mathrm{A}(0)$ is the atomic density matrix and $\rho_\mathrm{F}(0)$ is the field density matrix. The atom is assumed to be initially in its ground state, $\rho_\mathrm{A}(0) = \ketbra{g}{g}$.

To describe the initial state $\rho_\mathrm{F}(0)$ of the field, we first review some key concepts regarding quasi-probability distributions. In general, the density matrix for an arbitrary set of light modes $\vec{\nu}=(\nu_1, \ldots, \nu_k)$ of dimension $k$ can be written in terms of the multidimensional positive $P$ representation and projection operator $\Lambda(\vec{\alpha},\vec{\beta}^*)$~\cite{drummond1980}:
\begin{equation}
    \rho_{\vec{\nu}} = \int{\dd[2k]{\vec{\alpha}} \dd[2k]{\vec{\beta}} P_{\vec{\nu}}(\vec{\alpha}, \vec{\beta}^*)\Lambda_{\vec{\nu}}(\vec{\alpha}, \vec{\beta}^*)},
    \label{eq:density_projection}
\end{equation}
where the projection operator is $\Lambda_{\vec{\nu}}(\vec{\alpha}, \vec{\beta}^*) = \frac{\ketbra*{\vec{\alpha}}{\vec{\beta}}}{\braket*{\vec{\beta}}{\vec{\alpha}}}=\prod_{i=1}^k\frac{\ketbra{\alpha_i}{\beta_i}}{\braket{\alpha_i}{\beta_i}}$, the kets $\ket{\vec{\alpha}}=\prod_{i=1}^k\ket{\alpha_i}$ and $\ket*{\vec{\beta}}=\prod_{i=1}^k\ket{\beta_i}$ denote multimode coherent states with complex number parameters $\vec{\alpha}=(\alpha_1, \ldots, \alpha_k)$ and $\vec{\beta}=(\beta_1, \ldots, \beta_k)$, and the coherent state overlap matrix elements satisfy $\braket*{\vec{\beta}}{\vec{\alpha}} = e^{-\frac12 (|\vec{\alpha}|^2 + |\vec{\beta}|^2 - 2\vec{\alpha}^*\cdot\vec{\beta} )}$. The positive $P$ representation takes the form
\begin{align}
    P_{\vec{\nu}}(\vec{\alpha}, \vec{\beta}^*) &= \frac{1}{(2\pi)^{2k}}e^{\frac{-|\vec{\alpha} - \vec{\beta}|^2}{4}}\expval{\rho_{\vec{\nu}}}{\frac{\vec{\alpha} + \vec{\beta}}{2}} \nonumber\\
    &= \frac{1}{(4\pi)^{k}}e^{\frac{-|\vec{\alpha} - \vec{\beta}|^2}{4}} Q_{\vec{\nu}}\left(\frac{\vec{\alpha} + \vec{\beta}}{2}\right),
\end{align}
where $Q_{\vec{\nu}}(\vec{\alpha})=\frac{1}{\pi^k}\expval{\rho_{\vec{\nu}}}{\vec{\alpha}}$ is the Husimi distribution containing the quasi-probability distribution of light states in the multimode coherent state basis. Note that the bras and kets here are written without explicit mode indexes for compactness; they are to be understood as states in the Hilbert space of the set of modes $\vec{\nu}$ of interest.

Since only the modes associated with $\vec{\mathrm{p}}=(\mathrm{p}_1, \ldots, \mathrm{p}_k)$ are initially active and all other modes are in the vacuum state, the density matrix of the electromagnetic field at $t = 0$ can be written as
\begin{equation}
    \rho(0) = \int{\dd[2k]{\vec{\alpha}} \dd[2k]{\vec{\beta}} P(\vec{\alpha}, \vec{\beta}^*)\frac{\ketbra*{\vec{\alpha}}{\vec{\beta}}}{\braket*{\vec{\beta}}{\vec{\alpha}}}} \otimes \prod\limits_{\nu \notin \vec{\mathrm{p}}} \ketbra{0_{\nu}}{0_{\nu}},
    \label{eq:density_EM}
\end{equation}
where $\vec{\alpha}$ and $\vec{\beta}$ are implicitly assumed to correspond to the coherent amplitudes of the modes $\vec{\mathrm{p}}$ from now on.
Using the fact that the von Neumann equation \autoref{eq:density_equation} is linear, and that $P(\vec{\alpha}, \vec{\beta}^*)$ is independent of time, it is possible to define $\rho_{\vec{\alpha} \vec{\beta}}(t)$ such that it satisfies
\begin{equation}
    i\hbar  \pdv{\rho_{\vec{\alpha} \vec{\beta}}(t)}{t}  = [{H_o}(t), \rho_{\vec{\alpha} \vec{\beta}}(t)],
    \label{eq:rho_alphabeta(t)}
\end{equation}
where $\rho_{\vec{\alpha} \vec{\beta}}(0) = \ketbra{g}{g} \otimes \ketbra*{\vec{\alpha}}{\vec{\beta}} \otimes\prod\limits_{\nu \notin \vec{\mathrm{p}}} \ketbra{0_{\nu}}{0_{\nu}}$. The time-dependent version $\rho_{\vec{\alpha} \vec{\beta}}(t)$ is implicitly defined from the complete density matrix operator via
\begin{equation}
    \rho(t) = \int{\dd[2k]{\vec{\alpha}} \dd[2k]{\vec{\beta}} \frac{P(\vec{\alpha}, \vec{\beta}^*)}{\braket*{\vec{\beta}}{\vec{\alpha}}} \rho_{\vec{\alpha} \vec{\beta}}(t)}.
    \label{eq:rho(t)}
\end{equation}
Following steps analogous to those in Ref.~\cite{gorlach2023}, and given in more detail in \appref{app:schrodinger}, the problem can be rewritten such that it is only necessary to solve a Schrödinger equation:
\begin{subequations}
\label{eq:psi_alpha}
\begin{gather}
    i\hbar\pdv{\ket{\psi_{\vec{\alpha}}(t)}}{t} = H_{\vec{\alpha}}(t) \ket{\psi_{\vec{\alpha}}(t)},\\
    H_{\vec{\alpha}}(t) = H_\mathrm{A} + \vec{d}_\mathrm{A} \cdot (\vec{E}(\vec{r},t) + \Vec{E}_{\vec{\alpha}}(\vec{r},t))
\end{gather}
\end{subequations}
where the initial condition is $\ket{\psi_{\vec{\alpha}}(0)} = \ket{g} \otimes \prod_{\nu}\ket{0_{\nu}}$, and the new term contains a classical field given by $\vec{E}_{\vec{\alpha}}(t) = \sum_{i=1}^k\vec{g}_\mathrm{p_i}(\vec{r},t)\alpha_i + \text{c.c.}$ in addition to the electric field operator $\vec{E}(\vec{r},t)$. We note that the classical field is simply a vector-valued function of time and space, and there is no fundamental difference between a ``single-mode'' and ``multi-mode'' classical field---in other words, a classical field can always be written as a coherent state of a single temporal mode.

Using the same assumptions and approximations as in Refs.~\cite{gorlach2020, lewenstein2021}, the electromagnetic field state can be written as a product of coherent states. This corresponds to a mean-field approximation for the back-action of the atom on the electromagnetic field, or equivalently to including only the coherent fields produced by atomic dipole oscillations and neglecting incoherent contributions due to fluctuations. The coherent contributions dominate when spontaneous emission is not important, as is the case for the HHG spectrum. Under these approximations, $\ket{\psi_{\vec{\alpha}}(t)}$ may be written as a tensor product:
\begin{equation}
    \ket{\psi_{\vec{\alpha}}(t)} = \ket{\phi_{\vec{\alpha}}(t)} \!\otimes\! \prod\limits_{\ell}\ket{\gamma^{\vec{\alpha}}_{\ell}(t)} = \ket{\phi_{\vec{\alpha}}(t)} \!\otimes\! \ket{\vec{\gamma}_{\vec{\alpha}}(t)},
    \label{eq:psi_product}
\end{equation}
describing a separable and coherent output state of light characterized by a vector of coherent field amplitudes $\vec{\gamma}_{\vec{\alpha}}(t) = (\gamma^{\vec{\alpha}}_{\ell_1}(t), \gamma^{\vec{\alpha}}_{\ell_2}(t), \ldots)$, where for the single-mode amplitudes, $\vec{\alpha}$ is used as a superscript for compactness of notation. Within the dipolar mean-field approximation, nonclassical effects in the HHG spectrum, such as mode entanglement, squeezing or non-Gaussianity, are sure to be absent. Although this is generally expected to be a good approximation, recent works~\cite{lewenstein2021,Theidel2024,Theidel2025,Tzur2025} have demonstrated the possibility of observing beyond-mean-field effects in the output state of HHG\@. Moreover, we note that while the assumptions considered here lead to the separability of the light and matter degrees of freedom, recent works in strong-field ionization have also studied cases in which this does not hold~\cite{nandi2024,rivera-dean2025}.
As we are here interested not in the quantum correlations in the output light, but in the ``opposite'' question of how the presence of quantum correlations in the input light affects the overall HHG spectrum (which is dominated by the mean-field contribution), we do not consider these effects here.
Going back to \autoref{eq:psi_product}, the first term $\ket{\phi_{\vec{\alpha}}(t)}$ satisfies the Schrödinger equation for the atom and its interaction with the classical field:
\begin{equation}
    i\hbar\pdv{\ket{\phi_{\vec{\alpha}}(t)}}{t} = \left(H_\mathrm{A} + \Vec{d} \cdot \Vec{E}_{\vec{\alpha}}(t)\right)\ket{\phi_{\vec{\alpha}}(t)}.
    \label{eq:phi_alpha}
\end{equation}
The second term, $\ket{\vec{\gamma}_{\vec{\alpha}}(t)}$, represents the coherent states of light written in the original plane-wave basis where the frequency is well-defined.
The coherent state amplitudes can be expressed in terms of the Fourier transform of the expectation value of the atomic dipole moment operator
\begin{equation}
    \begin{aligned}
    &\gamma^{\vec{\alpha}}_{\ell}(t) = -\frac{1}{\hbar} \vec{d}_{\vec{\alpha}}(\omega_\ell,t) \cdot \vec{f}_{\ell}(\vec{0}) \\
    &\vec{d}_{\vec{\alpha}}(\omega,t) = \int^t_{-\infty}e^{i\omega\tau} \expval{\vec{d}}{\phi_{\vec{\alpha}}(\tau)}\dd{\tau}.
    \label{eq:gamma_coherent}
    \end{aligned}
\end{equation}
Thus, the dipole moment $\vec{d}_{\vec{\alpha}}(\omega,t)$ and consequently the coherent amplitudes $\vec{\gamma}_{\vec{\alpha}}(t)$ of the output state will depend parametrically on all the coherent amplitudes of the input modes contained in $\vec{\alpha}$. This dependence is in turn given by the solutions of the atomic part of the Schrödinger equation driven by the associated classical field. We are interested in the asymptotic state in the limit $t \to \infty$ when all HHG processes have concluded.
For simplicity of notation, we thus take this limit implicitly and drop most explicit time arguments in the following.

The density matrix operator is then given by
\begin{multline}
    \rho = \int{\dd[2k]{\vec{\alpha}} \dd[2k]{\vec{\beta}}\frac{P(\vec{\alpha}, \vec{\beta}^*)}{\braket*{\vec{\beta}}{\vec{\alpha}}} \ketbra*{\phi_{\vec{\alpha}}}{\phi_{\vec{\beta}}} } \otimes \\
    \times D(\vec{\alpha}) \ketbra*{\vec{\gamma}_{\vec{\alpha}}}{\vec{\gamma}_{\vec{\beta}}}D^{\dagger}(\vec{\beta}),
    \label{eq:density_matrix}
\end{multline}
where $D(\vec{\alpha}) = \prod_{i=1}^k D(\alpha_i) = e^{\vec{\alpha} \cdot \vec{b}^\dagger - \vec{\alpha}^* \cdot \vec{b}}$ is the multimode displacement operator for the pulse modes $\vec{p}$, with $\vec{b} = (b_1, \ldots, b_k)$ the vector of annihilation operators for the temporal modes.
We note that within this coherent-response approximation, the trace of the density matrix is not necessarily unity, i.e., the norm of the state is not preserved analytically.
We will show later that the deviation of the trace from unity is negligible in the single-atom case, but can become significant in the many-atom limit.
To apply this operator, we rewrite it in the plane-wave basis using the unitary transformation introduced above, which gives $D(\vec{\alpha}) = e^{\tilde{\vec{\alpha}} \cdot \vec{a}^\dagger - \tilde{\vec{\alpha}}^* \cdot \vec{a}}$, where $\tilde{\vec{\alpha}} = (\tilde{\alpha}_{\ell_1}, \tilde{\alpha}_{\ell_2}, \ldots)$ is a vector of coherent amplitudes in the plane-wave basis, with $\tilde{\vec{\alpha}} = \vec{U} \vec{\alpha}$.
Using that the action of a displacement operator on a coherent state is $D(\alpha)\ket{\beta} = e^{(\alpha\beta^* - \alpha^*\beta)/2} \ket{\alpha + \beta}$, we get
\begin{equation}
    D(\vec{\alpha}) \ket{\vec{\gamma}_{\vec{\alpha}}} = 
    e^{i\varphi_{\vec{\alpha}}} \ket{\vec{\gamma}_{\vec{\alpha}} + \tilde{\vec{\alpha}}},
    \label{eq:Gamma}
\end{equation}
where $\varphi_{\vec{\alpha}} = \frac{\tilde{\vec{\alpha}} \cdot \vec{\gamma}_{\vec{\alpha}}^* - \tilde{\vec{\alpha}}^* \cdot \vec{\gamma}_{\vec{\alpha}}}{2i}$.
\autoref{eq:density_matrix} then becomes
\begin{multline}
    \rho = \int \dd[2k]{\vec{\alpha}} \dd[2k]{\vec{\beta}}
    \frac{P(\vec{\alpha}, \vec{\beta}^*)}{\braket*{\vec{\beta}}{\vec{\alpha}}}
    \ketbra*{\phi_{\vec{\alpha}}}{\phi_{\vec{\beta}}} \otimes \\
    e^{i(\varphi_{\vec{\alpha}} - \varphi_{\vec{\beta}})} 
    \ketbra*{\vec{\gamma}_{\vec{\alpha}} + \tilde{\vec{\alpha}}}{\vec{\gamma}_{\vec{\beta}} + \tilde{\vec{\beta}}}
    \label{eq:rho_new2}
\end{multline}

\subsection{HHG spectrum}
Having obtained the density matrix describing the system, we can now calculate the HHG spectrum by considering the expectation value of the electromagnetic energy:
\begin{equation}
    \varepsilon = \sum\limits_\ell \Tr(\hbar \omega_\ell {a}_{\ell}^{\dagger}{a}_{\ell} \rho)
    = \sum\limits_\ell \varepsilon_\ell.
    \label{eq:energy}
\end{equation}
With $\rho$ from \autoref{eq:rho_new2}, $ \varepsilon_\ell$ becomes
\begin{multline}
    \varepsilon_\ell = \hbar \omega_\ell \int \dd[2k]{\vec{\alpha}} \dd[2k]{\vec{\beta}}
    \frac{P(\vec{\alpha}, \vec{\beta}^*)}{\braket*{\vec{\beta}}{\vec{\alpha}}}
    \braket*{\phi_{\vec{\beta}}}{\phi_{\vec{\alpha}}} \\
    \times e^{i(\varphi_{\vec{\alpha}} - \varphi_{\vec{\beta}})} 
    \mel*{\vec{\gamma}_{\vec{\beta}} + \tilde{\vec{\beta}}}{a_{\ell}^{\dagger} a_{\ell}}{\vec{\gamma}_{\vec{\alpha}} + \tilde{\vec{\alpha}}}.
    \label{eq:energy_quanta}
\end{multline}
As coherent states are eigenstates of the annihilation operator and each operator only acts on its corresponding mode, we can apply the annihilation and creation operators to the wave functions on the right and left, respectively, using that $a_{\ell} \ket{\vec{\gamma}_{\vec{\alpha}} + \tilde{\vec{\alpha}}} = (\gamma^{\vec{\alpha}}_{\ell} + \tilde{\alpha}_{\ell}) \ket{\vec{\gamma}_{\vec{\alpha}} + \tilde{\vec{\alpha}}}$.
Further inserting the previously mentioned expressions for the overlap for coherent states $\braket*{\vec{\beta}}{\vec{\alpha}}$, for the phases $\varphi^{\vec{\alpha}}_{\ell}$, and for $P(\vec{\alpha}, \vec{\beta}^*)$, \autoref{eq:energy_quanta} can be expanded as
\begin{multline}
    \varepsilon_\ell = \hbar \omega_\ell \int \dd[2k]{\vec{\alpha}} \dd[2k]{\vec{\beta}} \frac{e^{\frac{-|\vec{\alpha} - \vec{\beta}|^2}{4}}}{{(4\pi)^k}}Q\left(\frac{\vec{\alpha} + \vec{\beta}}{2}\right) \braket*{\phi_{\vec{\beta}}}{\phi_{\vec{\alpha}}} \\
    \times (\gamma^{\vec{\beta}}_{\ell} + \tilde{\beta}_{\ell})^* (\gamma^{\vec{\alpha}}_{\ell} + \tilde{\alpha}_{\ell}) \\
    \times \exp\Big[\vec{\gamma}_{\vec{\alpha}}^* \cdot \vec{\gamma}_{\vec{\beta}} - \frac12 |\vec{\gamma}_{\vec{\alpha}}|^2 - \frac12 |\vec{\gamma}_{\vec{\beta}}|^2 \\
    + \tilde{\vec{\alpha}}^* \cdot (\vec{\gamma}_{\vec{\beta}} - \vec{\gamma}_{\vec{\alpha}}) - \tilde{\vec{\beta}} \cdot (\vec{\gamma}_{\vec{\beta}}^* - \vec{\gamma}_{\vec{\alpha}}^*) \Big].
    \label{eq:Energy_reworked}
\end{multline}
Our interest lies in the high-harmonic spectrum, i.e., in $\epsilon_\ell$ for indices $\ell$ with $\omega_\ell \gg \omega_0$.
As we want to study cases where the driving pulse generates high harmonics, but does not contain them already, we can neglect the coherent driving amplitudes relative to the pulse-generated amplitudes, $\gamma^{\vec{\alpha}}_{\ell} + \tilde{\alpha}_{\ell} \approx \gamma^{\vec{\alpha}}_{\ell}$, and similarly for $\vec{\alpha} \to \vec{\beta}$. Note that this only applies to the prefactor, not to the dot products in the exponential which contain sums over all modes, including the driving modes.
Furthermore, we perform a change of variables
\begin{equation}
    \begin{cases}
        \vec{\alpha}_m = \frac12(\vec{\alpha} + \vec{\beta}), \\
        \delta\vec{\alpha} \; = \frac12(\vec{\alpha} - \vec{\beta}),
    \end{cases}
    \label{eq:Change_alpha}
\end{equation}
and introduce $\delta\vec{\gamma} = \vec{\gamma}_{\vec{\alpha}} - \vec{\gamma}_{\vec{\beta}}$, which lets us rewrite the expression for the energy spectrum as
\begin{multline}
    \varepsilon_\ell =  \hbar \omega_\ell \int \dd[2k]{\vec{\alpha}_m} \dd[2k]{\delta\vec{\alpha}}
        \frac{e^{{-|\delta\vec{\alpha}|^2}}}{\pi^k} Q(\vec{\alpha}_m)
        \braket*{\phi_{\vec{\beta}}}{\phi_{\vec{\alpha}}} \\
    \times \gamma^{\vec{\alpha}}_{\ell}{\gamma^{\vec{\beta}*}_{\ell}} 
    e^{i\Im(\vec{\gamma}_{\vec{\alpha}}^* \cdot \vec{\gamma}_{\vec{\beta}}) - \frac12 |\delta\vec{\gamma}|^2 - \tilde{\vec{\alpha}}^* \cdot \delta\vec{\gamma} + \tilde{\vec{\beta}} \cdot \delta\vec{\gamma}^*},
    \label{eq:Energy_reworked2}
\end{multline}
where $\vec{\alpha} = \vec{\alpha}_m + \delta\vec{\alpha}$ and $\vec{\beta} = \vec{\alpha}_m - \delta\vec{\alpha}$ as well as $\tilde{\vec{\alpha}} = \vec{U} \vec{\alpha}$ and $\tilde{\vec{\beta}} = \vec{U} \vec{\beta}$ are to be understood as implicit functions of the new variables.

Instead of the mode-resolved spectrum $\varepsilon_\ell$, we are typically interested in the total frequency-resolved spectrum $\dv{\varepsilon}{\omega}$.
We obtain this by taking the free-space limit $V\to\infty$, corresponding to $\sum_{\ell} \rightarrow \frac{V}{(2\pi)^3c^3} \int_0^{\infty} \dd{\omega}\omega^2 \sum_{\sigma}\int \dd{\Omega}$, and grouping all terms with the same frequency, which corresponds to an angular integral over the solid angle $\Omega$ and a sum over the two polarization states $\sigma$. Using \autoref{eq:gamma_coherent} to express the coherent amplitudes in terms of the dipole moment, the polarization sum and angular integral can be performed analytically, yielding
\begin{multline}
    \dv{\varepsilon}{\omega} = \frac{\omega^4}{6\pi^2 \epsilon_0 c^3} \int \dd[2k]{\vec{\alpha}_m} \dd[2k]{\delta\vec{\alpha}}
        \frac{e^{{-|\delta\vec{\alpha}|^2}}}{\pi^k} Q(\vec{\alpha}_m)
        \braket*{\phi_{\vec{\beta}}}{\phi_{\vec{\alpha}}} \\
    \times \vec{d}_{\vec{\alpha}}(\omega) \cdot \vec{d}^*_{\vec{\beta}}(\omega)
    \, e^{i\Im(\vec{\gamma}_{\vec{\alpha}}^* \cdot \vec{\gamma}_{\vec{\beta}}) - \frac12 |\delta\vec{\gamma}|^2 - \tilde{\vec{\alpha}}^* \cdot \delta\vec{\gamma} + \tilde{\vec{\beta}} \cdot \delta\vec{\gamma}^*},
    \label{eq:spectrum_freqdiff}
\end{multline}
Since the integrand contains a Gaussian in $\delta\vec{\alpha}$, the main contributions to the integral will come from values of $\delta\vec{\alpha}$ close to the zero.
At the same time, HHG only occurs in intense pulses, which typically corresponds to large photon numbers, implying $|\vec{\alpha}_m| \gg |\delta\vec{\alpha}|$.
Note that this argument is independent of the choice of (temporal) mode basis used to describe the driving light, as $|\vec{\alpha}_m|$ is invariant under unitary transformations.
Expanding all terms except for the Gaussian to lowest order in $\delta\vec{\alpha}$ then corresponds to a multimode ``diagonal'' approximation $\vec{\alpha} \approx \vec{\beta} \approx \vec{\alpha}_m$, and the remaining Gaussian can be integrated analytically. The argument presented here to justify the diagonal approximation is thus independent of taking the limit of infinite quantization volume. Within this approximation, the energy spectrum simplifies to
\begin{equation}
    \dv{\varepsilon}{\omega}^\mathrm{diag} = \frac{\omega^4}{6\pi^2 \epsilon_0 c^3} \int \dd[2k]\vec{\alpha} Q(\vec{\alpha}) |\vec{d}_{\vec{\alpha}}(\omega)|^2.
    \label{eq:spectrum_gorlach2}
\end{equation}
In the last step, we used that the atom wavefunction $\ket{\phi_{\vec{\alpha}}}$ is normalized.
This result extends the single-plane-wave-mode result of Ref.~\cite{gorlach2023} to the case of arbitrary multimode driving fields without relying on any arguments related to the quantization volume. It has the same overall form: A weighted average of the dipole spectrum $|d_{\vec{\alpha}}(\omega)|^2$ over the (multimode $2k$-dimensional) Husimi distribution $Q(\vec{\alpha})$ of the initial state of the driving light.
We next study the accuracy of the diagonal approximation. Grouping all the terms that do not depend on $\delta\vec{\alpha}$ outside the respective integral from \autoref{eq:Energy_reworked2}, the emission spectrum can be written as
\begin{equation}
    \dv{\varepsilon}{\omega} = \frac{\omega^4}{6\pi^2 \epsilon_0 c^3} \int \dd[2k]{\vec{\alpha}_m} C_\omega(\vec{\alpha}_m)
        Q(\vec{\alpha}_m) |\vec{d}_{\vec{\alpha}}(\omega)|^2
    \label{eq:Energy_correction}
\end{equation}
where $C_\omega(\vec{\alpha}_m)$ is a correction factor that does not depend on the state of the light pulse and is given by
\begin{multline}
    C_\omega(\vec{\alpha}_m) = \int \dd[2k]{\delta\vec{\alpha}}
        \frac{e^{{-|\delta\vec{\alpha}|^2}}}{\pi^k}
        \frac{\vec{d}_{\vec{\alpha}}(\omega) \cdot \vec{d}^*_{\vec{\beta}}(\omega)}{|\vec{d}_{\vec{\alpha}_m}(\omega)|^2}
        \braket*{\phi_{\vec{\beta}}}{\phi_{\vec{\alpha}}} \\
        \times e^{i\Im(\vec{\gamma}_{\vec{\alpha}}^* \cdot \vec{\gamma}_{\vec{\beta}}) - \frac12 |\delta\vec{\gamma}|^2 - \tilde{\vec{\alpha}}^* \cdot \delta\vec{\gamma} + \tilde{\vec{\beta}} \cdot \delta\vec{\gamma}^*},
    \label{eq:Correction_factor}
\end{multline}
The validity of the diagonal approximation then depends on how much $C_\omega(\vec{\alpha}_m)$ differs from unity for the values of $\vec{\alpha}_m$ that contribute significantly to the Husimi distribution $Q(\vec{\alpha}_m)$ for a given driving field. To calculate this correction factor explicitly, we express $\vec{\gamma}$, $\vec{\alpha}$, and $\vec{\beta}$ inside the exponential in terms of the physical dipole moments and electric fields. The continuous limit for the first part gives
\begin{multline}
    t_1(\vec{\alpha},\vec{\beta}) = i\Im(\vec{\gamma}_{\vec{\alpha}}^* \cdot \vec{\gamma}_{\vec{\beta}}) - \frac12 |\delta\vec{\gamma}|^2 = 
    \frac{1}{12\pi^2 \hbar c^3 \epsilon_0} \\ 
    \int_0^{\infty} \dd{\omega}\omega^3 
    \left(2i\Im[\vec{d}^*_{\vec{\alpha}}(\omega) \cdot \vec{d}_{\vec{\beta}}(\omega)] - |\delta \vec{d}(\omega)|^2\right)
    \label{eq:First_term_exponential}
\end{multline}
where we have introduced $\delta \vec{d}(\omega) = \vec{d}_{\vec{\alpha}}(\omega) - \vec{d}_{\vec{\beta}}(\omega)$.
The second part of the exponential can be rewritten in terms of the Fourier transform of the driving electric field (see \appref{app:E_field_transform} for details), giving
\begin{multline}
    t_2(\vec{\alpha},\vec{\beta}) = - \tilde{\vec{\alpha}}^* \cdot \delta\vec{\gamma} + \tilde{\vec{\beta}} \cdot \delta\vec{\gamma}^* = \\
    \frac{1}{\sqrt{2\pi} \hbar} \int\limits_0^{\infty} \dd{\omega} \left[\Vec{E}^{+}_{\vec{\beta}}(\omega) \cdot \delta\vec{d}^*(\omega) -\Vec{E}^{+*}_{\vec{\alpha}}(\omega) \cdot \delta\vec{d}(\omega) \right],
    \label{eq:Second_term_reworked}
\end{multline}
where $\Vec{E}^{+}_{\vec{\alpha}}(\omega)$ is the positive-frequency part of the Fourier transform of the driving electric field  with coherent amplitudes $\vec{\alpha}$. This form makes clear that the correction due to $t_2(\vec{\alpha},\vec{\beta})$ accounts for components of the generated field that overlap spectrally with the driving field.

The complete expression for the correction factor can then be written as
\begin{subequations}
\begin{align}
C_\omega(\vec{\alpha}_m) &= \int \dd[2k]{\delta\vec{\alpha}} \frac{e^{-|\delta\vec{\alpha}|^2}}{\pi^k}
        f_{\omega}(\vec{\alpha}_m,\delta\vec{\alpha}),\\
f_{\omega}(\vec{\alpha}_m,\delta\vec{\alpha}) &= \frac{\vec{d}_{\vec{\alpha}}(\omega) \cdot \vec{d}^*_{\vec{\beta}}(\omega)}{|\vec{d}_{\vec{\alpha}_m}(\omega)|^2} f_{\text{ov}}(\vec{\alpha}_m,\delta\vec{\alpha}),\\
f_{\text{ov}}(\vec{\alpha}_m,\delta\vec{\alpha}) &= \braket*{\phi_{\vec{\beta}}}{\phi_{\vec{\alpha}}} e^{t_1(\vec{\alpha},\vec{\beta}) + t_2(\vec{\alpha},\vec{\beta})},
\end{align}
\end{subequations}
where $f_{\text{ov}}(\vec{\alpha}_m,\delta\vec{\alpha})$ does not depend on the emission frequency $\omega$, and $\vec{\alpha} = \vec{\alpha}_m + \delta\vec{\alpha}$ and $\vec{\beta} = \vec{\alpha}_m - \delta\vec{\alpha}$ are implicit functions of $\vec{\alpha}_m$ and $\delta\vec{\alpha}$. The diagonal approximation is equivalent to assuming $f_\omega(\vec{\alpha}_m,\delta\vec{\alpha}) \approx 1$.

We remark that the final form for $f_{\omega}(\vec{\alpha}_m,\delta\vec{\alpha})$ does not depend explicitly on the coherent amplitudes $\vec{\alpha}_m$ and $\delta\vec{\alpha}$ themselves, but only on the corresponding classical fields $\Vec{E}^{+}_{\vec{\alpha}}(\omega)$ and $\Vec{E}^{+}_{\vec{\beta}}(\omega)$, as the atomic dipole expectation values are also fully determined by these fields.

Motivated by the fact that $f_{\text{ov}}(\vec{\alpha}_m,\delta\vec{\alpha})$ does not depend on the emission frequency $\omega$, we now check unitarity, i.e., the conservation of the trace of the density matrix. As mentioned above, the trace is explicitly conserved in the derivation until the coherent-response approximation is made, but it is not guaranteed to be conserved after this approximation. Starting from \autoref{eq:density_matrix} and performing similar steps as above, we obtain
\begin{gather}
    \Tr(\rho) = \int\dd[2k]{\vec{\alpha}_m} C_{\rho}(\vec{\alpha}_m) Q(\vec{\alpha}_m)\\
    C_{\rho}(\vec{\alpha}_m) = \int\dd[2k]{\delta\vec{\alpha}} \frac{e^{-|\delta\vec{\alpha}|^2}}{\pi^k} f_{\text{ov}}(\vec{\alpha}_m,\delta\vec{\alpha}),
    \label{eq:c_rho}
\end{gather}
where $C_{\rho}(\vec{\alpha}_m) = 1$ if $f_{\text{ov}}(\vec{\alpha}_m,\delta\vec{\alpha}) = 1$, which then gives $\Tr(\rho) = 1$.
This thus reveals that any deviation of the HHG-spectrum correction factor $C_\omega(\vec{\alpha}_m)$ from unity due to $f_{\text{ov}}(\vec{\alpha}_m,\delta\vec{\alpha})$ is actually due to the non-unitarity of the coherent-response approximation, and thus does not necessarily imply a failure of the diagonal approximation.
In contrast, a deviation of the dipole correction factor $\vec{d}_{\vec{\alpha}}(\omega) \cdot \vec{d}^*_{\vec{\beta}}(\omega) / |\vec{d}_{\vec{\alpha}_m}\!(\omega)|^2$ from unity is a direct consequence of the diagonal approximation, and thus indeed indicates a failure of this approximation.
We note that in the diagonal approximation, where $C_{\rho}(\vec{\alpha}_m) = C_\omega(\vec{\alpha}_m) = 1$, the trace of the density matrix is conserved, such that this second approximation ``cures'' the non-unitarity of the coherent-response approximation.

\subsection{Many-atom case}
\label{sec:many-atom}
Up to now, we have only considered the response of a single atom. However, in realistic experimental scenarios, the driving laser excites an ensemble of emitters.
We will now consider the case of HHG for $N$ identical, non-interacting, and phase-matched (i.e., closely spaced) atoms.
The only terms in the full Hamiltonian of \autoref{eq:Hamiltonian} that will be affected are the atomic part, which is extended to include $N$ atoms as $H_A = \sum_{i=1}^N \frac{\vec{p}^2_i}{2m}+U_i(\vec{r}_i)$, and the dipole operator, which is extended in the same fashion as $\Vec{d}_A = \sum_{i=1}^N\Vec{d}_{A,i}$. Accordingly, the atomic wavefunction becomes a many-body product state $\ket{\phi_{\vec{\alpha}}(t)} = \prod_{i=1}^N \ket{\phi_{\vec{\alpha},i}(t)}$, which is still a normalized wave function. As for the dipole moment operator, from \autoref{eq:gamma_coherent} the inclusion of multiple emitters will thus affect the output coherent state amplitudes. The Fourier transform of the dipole expectation value is then
\begin{multline}
    \vec{d}_{\vec{\alpha}}(\omega) = \int^t_{-\infty}e^{i\omega\tau} \prod_{i=1}^N \expval{\sum_{j=1}^N \vec{d}_j}{\phi_{\vec{\alpha},i}(t)}\dd{\tau}\\ = N \int^t_{-\infty}e^{i\omega\tau} \expval{\vec{d}_i}{\phi_{\vec{\alpha},i}(t)}\dd{\tau},
\end{multline}
where we used the normalization of the atomic states and the indistinguishability of the atoms. Note that the expectation value for the $i$-th atom that appears in the last equation is equal to the single-atom case, as within the approximations we employ, each atom is simply driven by the same external field and back-action is neglected. Therefore, the squared dipole term in the HHG spectrum of \autoref{eq:spectrum_gorlach2} will present the well-known $N^2$ enhancement factor due to $N$ coherent in-phase dipoles.

We now turn our attention to the many-atom correction factor to the diagonal approximation.
Since the dipole correction factor $\vec{d}_{\vec{\alpha}}(\omega) \cdot \vec{d}^*_{\vec{\beta}}(\omega) / |\vec{d}_{\vec{\alpha}_m}\!(\omega)|^2$ is a ratio of dipole operators, the factor $N$ cancels out, so that this term is independent of the number of atoms.
Thus, the only term that will be affected is $f_{\text{ov}}(\vec{\alpha}_m,\delta\vec{\alpha})$.
The atomic wavefunction overlap term becomes
\begin{equation}
    \braket*{\phi_{\vec{\beta}}}{\phi_{\vec{\alpha}}} = \prod_{i=1}^N \braket*{\phi_{\vec{\beta},i}}{\phi_{\vec{\alpha},i}} = \braket*{\phi_{\vec{\beta},i}}{\phi_{\vec{\alpha},i}}^N.
    \label{eq:fov_many_atom}
\end{equation}
Moreover, an enhancement factor of $N^2$ will appear in the first exponential term of \autoref{eq:First_term_exponential}, again due to the squared dipoles. On the other hand, the second exponential term of \autoref{eq:Second_term_reworked} depends linearly on the Fourier transform of the dipole expectation value, such that in the many-atom case it will only scale linearly with $N$. Then, the mode-independent correction term will be
\begin{equation}
    f_{\text{ov}}(\vec{\alpha}_m,\delta\vec{\alpha}) = \braket*{\phi_{\vec{\beta},i}}{\phi_{\vec{\alpha},i}}^N e^{N^2t_1(\vec{\alpha},\vec{\beta}) + Nt_2(\vec{\alpha},\vec{\beta})},
\end{equation}
where $t_1(\vec{\alpha},\vec{\beta})$ and $t_2(\vec{\alpha},\vec{\beta})$ are the single-atom exponential terms. 
We thus find that the many-atom overlap correction factor scales exponentially with the atomic number $N$ and with its square $N^2$.
However, this scaling is confined to the overlap correction factor $f_{\text{ov}}(\vec{\alpha}_m,\delta\vec{\alpha})$, which as discussed above indicates a breakdown of unitarity within the mean-field coherent-response approximation, and not of the diagonal approximation itself. The dipole correction factor $\vec{d}_{\vec{\alpha}}(\omega) \cdot \vec{d}^*_{\vec{\beta}}(\omega) / |\vec{d}_{\vec{\alpha}_m}\!(\omega)|^2$ is independent of $N$, and thus the validity of the diagonal approximation is not affected by the number of atoms.

\section{Results}
\label{sec.results}
\begin{figure}[tb]
   \includegraphics[width=\linewidth]{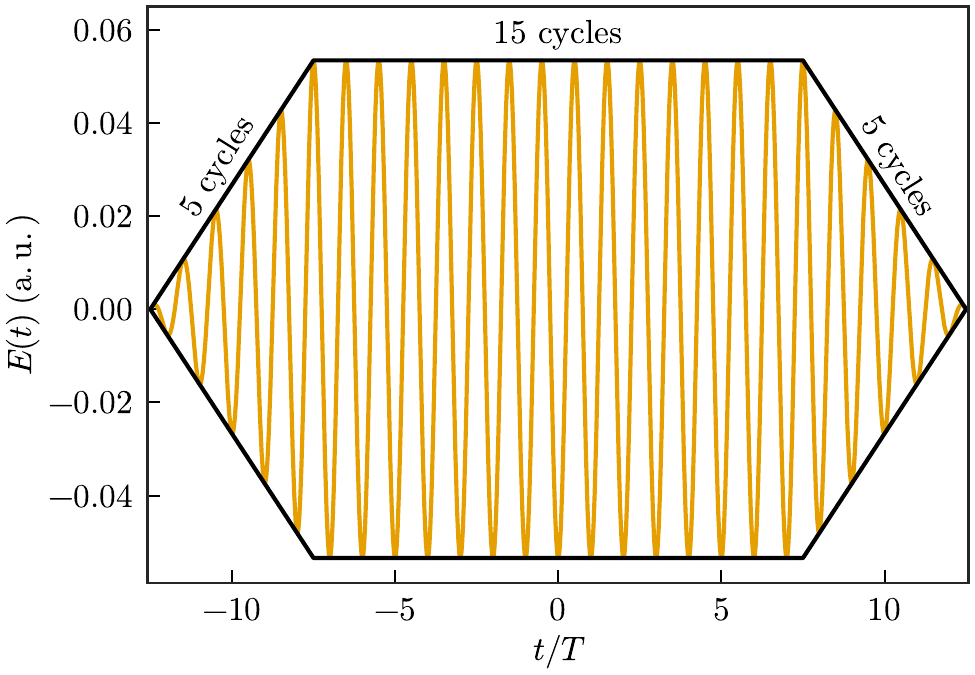}
   \caption{Illustration of the electric field pulse shape $E(t)$ used, given by a single-frequency carrier wave modulated with a flat-top pulse envelope. The envelope rises linearly over 5 cycles, becomes constant for 15 cycles, and then decreases linearly over 5 cycles. $E(t)$ is given in atomic units for a peak intensity of $I_0 = 10^{14}$~W/cm$^2$ and $t$ in units of the period $T = \frac{2\pi}{\omega_0}$.
   }
   \label{fig:Electric_field}
\end{figure}

In this section, we apply the developed theory to calculate the HHG spectrum.
We do this for a single temporal mode instead of for a multimode case, for various reasons:
First, as mentioned above, pulses of light with a controlled quantum state are typically generated in a single temporal mode, and thus the single-mode case is of particular interest~\cite{Rasputnyi2024, Demontigny2026}.
In such setups, even if the single-mode regime is not achieved, the temporal modes usually do not have controlled quantum correlations, but are rather in a statistical mixture, such that the HHG would would just be a weighted average of the single-mode results.
Third, in genuine multimode setups, even just the use of classical fields opens a huge parameter space, as the spatial and temporal shape, carrier frequency, polarization, and propagation direction of each mode can be chosen independently.
These freedoms are routinely exploited in classical HHG setups~\cite{Chini2014, Rego2019}, and in the quantum regime, there could be arbitrary quantum correlations between the modes, prohibiting a systematic general study of the multimode case.

As such, we now drop all vectorial notation used for the general multimode formalism, and assume a single temporal mode with index $\mathrm{p}$, such that $\Vec{E}_d(\vec{r},t) = \vec{g}_{\mathrm{p}}(\vec{r},t) {b}_{\mathrm{p}} + \mathrm{H.c.}$.
We analyze the correction factor $C_\omega(\alpha_m)$ to evaluate the validity of the diagonal approximation.
To facilitate comparison, we assume that the temporal profile of the pulse mode at the atom position is the same flat-top laser pulse as used in the numerical simulations in Ref.~\cite{gorlach2023}, given by
\begin{subequations}\label{eq:Electric_field}
\begin{align}
    &\vec{g}_\mathrm{p}(\vec{r}=0, t) = \hat{z} E_{1,\mathrm{p}} F_\text{flat}(t) e^{-i \omega_0 t}\\
    &F_\text{flat}(t) = \begin{cases}
        1                         & |t| \leq \frac{1}{2}T_f \\
        1 - \frac{|t|-\frac{1}{2}T_f}{T_r} & \frac{1}{2}T_f < |t| \leq T_r + \frac{1}{2}T_f \\
        0                         & |t| > T_r + \frac{1}{2}T_f
    \end{cases}
\end{align}
\end{subequations}
where $F_\text{flat}(t)$ is the linear flat-top pulse envelope, the carrier wavelength is $\lambda_0 = 800$~nm, corresponding to a period $T = \lambda_0/c$ and central frequency $\omega_0 = 2\pi/T$, and $T_r = 5 T$ and $T_f = 15 T$ are the durations of the linear ramp-on and flat parts of the pulse, respectively. This flat-top pulse is represented in \autoref{fig:Electric_field}. 

To ensure the $b_{\mathrm{p}}$ and $b^{\dagger}_{\mathrm{p}}$ operators follow the commutation relations described previously, the electric field must be properly normalized. To accomplish this, the full spatial shape of the mode function $\vec{g}_\mathrm{p}(\vec{r},t)$ needs to be defined, not just its value at the position of the atom. We assume that it corresponds to a laser pulse propagating in free space along the $x$-axis with a flat transverse profile with cross-sectional area $A=1~\mu$m$^2$, corresponding to a relatively well-focused beam. The temporal shape given by the flat-top envelope described above then also determines the spatial profile along the $x$-axis. The normalization procedure, detailed in \appref{app:normalization}, then gives
\begin{equation}
    E_{1,\mathrm{p}} = \sqrt{\frac{\hbar \omega_0}{2 \varepsilon_0 V_{\text{eff}}}},
    \label{eq:Electric1_field}
\end{equation}
where $V_{\text{eff}} \approx 18.34 \lambda_0 A$ is the effective pulse quantization volume for the pulse parameters above. The (temporal) peak intensity $I$ of the pulse is then given by
\begin{equation}
    I = 2 c \varepsilon_0 E_{1,\mathrm{p}}^2 \expval{b_\mathrm{p}^\dagger b_\mathrm{p}} = I_{1,\mathrm{p}} \expval{b_\mathrm{p}^\dagger b_\mathrm{p}} = I_{1,\mathrm{p}} |\alpha_\mathrm{p}|^2,
\end{equation}
where $I_{1,\mathrm{p}} = c \hbar \omega_0 / V_{\text{eff}}$ is the peak intensity per photon in the pulse mode, $\expval{b_\mathrm{p}^\dagger b_\mathrm{p}}$ is the photon number operator, and the equality $\expval{b_\mathrm{p}^\dagger b_\mathrm{p}} = |\alpha_\mathrm{p}|^2$ applies for coherent states. For the pulse parameters used here, we obtain values of $E_{1,\mathrm{p}} \approx 30{,}911$~V/m and $I_{1,\mathrm{p}} \approx 507.26$~W/cm$^2$. This means that peak intensities of $I_\mathrm{HHG} \approx 10^{14}$~W/cm$^2$ as required for HHG will only occur if the Husimi distribution $Q(\alpha_m)$ contains non-negligible contributions at $|\alpha_m|^2 \approx I_\mathrm{HHG} / I_{1,\mathrm{p}} \approx 2 \times 10^{11}$, corresponding to $|\alpha_m| \approx 4.4 \times 10^5$.

The large photon numbers needed for HHG imply that the diagonal approximation is expected to be accurate, as only values of $\delta\alpha$ close to unity are expected to contribute significantly to the integral in $C_\omega(\alpha_m)$ due to the suppression by the Gaussian factor $e^{-|\delta\alpha|^2}$, so that $\delta\alpha /  \alpha_m \sim 10^{-5} \ll 1$ holds for all relevant contributions. The expression for $f_\omega(\alpha_m,\delta\alpha)$ then suggests that its value will be close to unity, implying $C_\omega(\alpha_m) \approx 1$, such that the full energy spectrum, \autoref{eq:Energy_correction}, reduces to the diagonal approximation, \autoref{eq:spectrum_gorlach2}.

\subsection{Single-atom correction}\label{sec:single-atom-results}
\begin{figure}
    \centering
    \includegraphics[width=\linewidth]{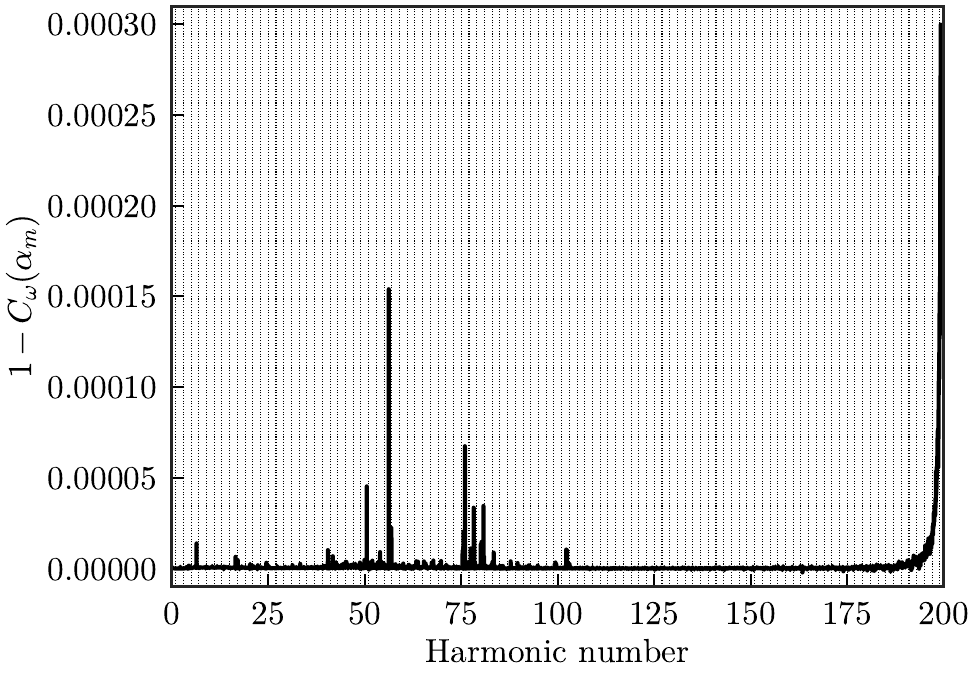}
    \caption{Evaluation of the difference $1 - C_\omega(\alpha_m)$ as a function of $\omega$. Vertical bands are a consequence of numerical noise.}
    \label{fig:cl_single}
\end{figure}

Next, with the definitions given above we will quantify the accuracy of the diagonal approximation by evaluating the integral correction factor $C_\omega(\alpha_m)$ for a single atom. We do so by solving the single-atom Schrödinger equation described in \autoref{eq:phi_alpha} and computing the expectation value of the dipole acceleration $\expval*{\ddot{d}}(t) = -\expval*{\dv{V}{z}} + E(t)$ and its Fourier transform. To maintain easy comparison with Ref.~\cite{gorlach2023}, we solve the dynamics using the same 1D model atom characterized by a soft Coulomb potential with the same ionization potential $I_p$ as neon:
\begin{align}
    H_A &= -\frac{1}{2}\dv[2]{}{z} + V(z); & V(z) &= -\frac{1}{\sqrt{z^2 + a^2}},
    \label{eq:Hamiltonian_atom}
\end{align}    
where $a = 0.8160~r_B$ is chosen such that $I_p = 0.7924$~Hartree. We set the mean coherent amplitude to $\alpha_m = \sqrt{I_\mathrm{HHG} / I_{1,\mathrm{p}}}$.  
The calculations were performed for various values of the real and imaginary parts of $\delta\alpha$, forming a grid in the complex plane in which the Gaussian integral was solved.
The full correction factor $C_\omega(\alpha_m)$ is shown in \autoref{fig:cl_single}.
As expected, the deviation remains small within the entire range considered, with values below $10^{-3}$.
Furthermore, the largest corrections, seen as narrow vertical bands in the figure, are mostly due to numerical instability near minima of the Fourier transform of the dipole expectation value $|\vec{d}_{\alpha_m}(\omega)|^2$.

These results indicate that the diagonal approximation $C_\omega(\alpha_m) \approx 1$ is indeed justified for the values of $\alpha_m$ relevant for HHG production within the integral over $Q(\alpha_m)$. 
This means that the diagonal approximation is accurate for obtaining the HHG spectrum with pulses in arbitrary quantum states in free space, consistent with the results in Ref.~\cite{gorlach2023}.
For the diagonal approximation to break down in the few-atom limit, one would need to consider situations where much smaller values of $\alpha_m$ are relevant, such that small deviations $\delta\alpha$ can have a significant effect.
This observation reinforces the general idea that many quantum effects in light-matter interaction require the photon number to be small, such that the quantized nature of light is relevant.
While this is not possible for HHG in free space, where the effective pulse volume is on the order of $\lambda_0^3$, it could potentially occur in nanophotonic environments where the mode volume is strongly reduced~\cite{liu2018,zograf2022,heimerl2025}, such that the same peak intensity can be reached with much smaller photon numbers.
For $\lambda_0=800$~nm, the number of photons required to reach $I_\mathrm{HHG} = 10^{14}$~W/cm$^2$ is given by $I_\mathrm{HHG} / I_{1,\mathrm{p}} = 13.4\,V[\mathrm{nm^3}]$.
This implies that for the smallest mode volumes predicted in plasmonic nanogap picocavities, on the order of $V \sim 1$~nm$^3$~\cite{Li2021Bright,Wu2021}, few-photon HHG could potentially be within the reach of experiment, and quantum effects could become significant.
Note that the quantum effects discussed here which could lead to a breakdown of the diagonal approximation only concern changes in the HHG spectrum that would not be reproduced by a statistical average over a collection of classical fields weighted by the Husimi function. 
In particular, we are not considering the quantum correlations in the generated HHG field, which are known to generated even with classical driving fields, and do not make any statement about whether such correlations could be affected by quantum correlations in the driving pulse. This would require a treatment going beyond the mean-field approximation in the generated field, which is beyond the scope of this work. 

\begin{figure}[tb]
   \includegraphics[width= \linewidth]{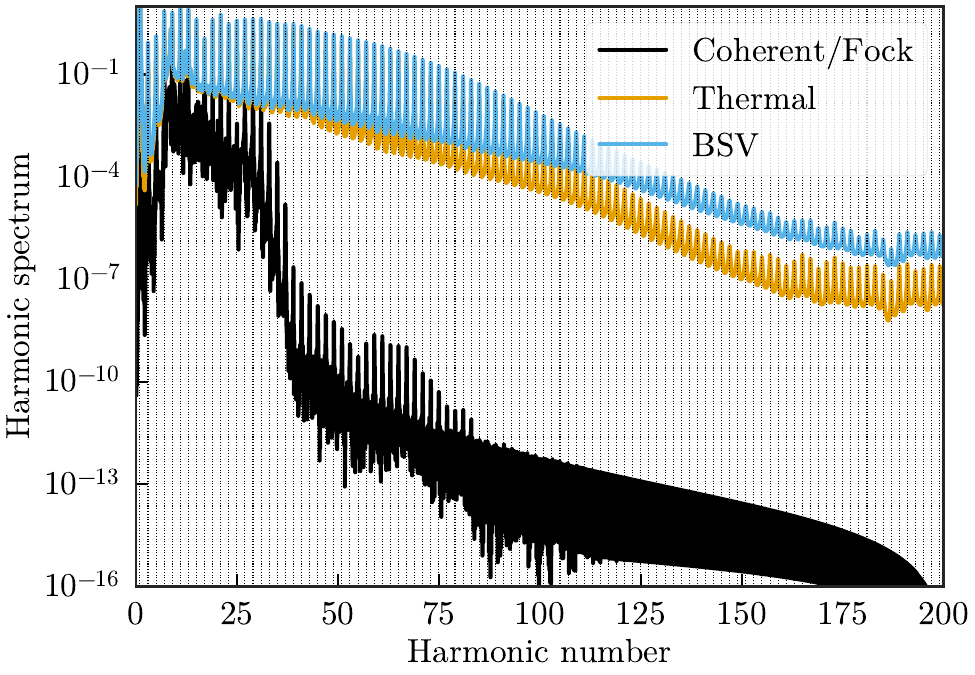}
   \caption{Harmonic spectrum of the different light distributions for a constant mean intensity. The peaks at odd multiples of the driving frequency $\omega_0$ arise because the electric field $E_{\alpha}(t)$ reaches a maximum (in absolute value) every half cycle.}
   \label{fig:Harmonic_comparison}
\end{figure}

For completeness, we next calculate the high harmonic spectrum for the same quantum light states considered in Ref.~\cite{gorlach2023}: coherent, Fock, thermal, and bright squeezed vacuum (BSV), represented by their respective Husimi distributions~\cite{kim1989properties}.
The same mean intensity, $I = 10^{14}$~W/cm$^2$, is chosen for each of these light states.
The HHG spectrum is shown in \autoref{fig:Harmonic_comparison}, where the amplitude of the response is plotted as a function of frequency (in terms of the harmonic number $\omega/\omega_0$).
Since the correction factor $C_\omega(\alpha_m)$ is very close to unity for all relevant $\alpha_m$, the results are indistinguishable from those obtained using the single-mode approximation of Ref.~\cite{gorlach2023}.
We mention that our numerical results differ slightly from those in Ref.~\cite{gorlach2023}, with more clearly visible peaks up to larger harmonic numbers.
This is most likely due to our use of a smooth cutoff of the dipole acceleration in time after the end of the pulse, which reduces numerical artifacts in the Fourier transform due to dipole oscillations between bound states that otherwise persist after the end of the pulse (infinitely if spontaneous emission is not taken into account).

The resulting spectra are indistinguishable for Fock and coherent states, since their Husimi distributions are very narrowly peaked around the mean intensity.
In contrast, thermal and BSV states give significantly higher HHG yields due to their much broader intensity distributions.
Based on the resulting emission spectrum, using thermal or BSV states instead of coherent or Fock states may appear more advantageous for generating high-harmonic pulses.

However, these differences do not represent genuine quantum effects.
While in principle the Husimi distribution is unique (i.e., the full quantum state can be reconstructed from it, which in turn means that two Husimi distributions are only identical if they describe the same exact density matrix), the fact that it is a real and positive quantity means that any sufficiently broad distribution can be reasonably approximated by an incoherent sum over unit-width Gaussians (i.e., coherent states).
Such a Husimi function corresponds to a statistical mixture of coherent (i.e., classical) states and thus cannot show any quantum effects in the emitted spectra from field excitation.
In other words, the same statistical properties of the light field can be obtained from an average over purely classical fields.
Note that the coherent state and thermal distributions already have this property even without any averaging over Gaussians. In contrast, BSV and Fock states are \emph{a priori} ``quantum'' states.

Even so, a corollary of our result regarding the near-unit value of the correction factor $C_\omega(\alpha)$ is the explicit confirmation that convolution of the Husimi function with a Gaussian of unit width in $\alpha$ will not significantly change the HHG spectrum for situations where many photons are required to reach intensities where HHG takes place.
This is because the validity of the diagonal approximation implies that the HHG spectrum changes little for variations of $\alpha$ on the order of unity (note that this is exactly the argument behind the conventional use of the semi-classical approximation in strong-field physics due to the well-known fact that fluctuations in coherent states are negligible for large photon numbers).
Nonetheless, such a convolution corresponds to a density matrix that is a classical statistical mixture of coherent states, $\rho = \int \dd[2]{\alpha} Q(\alpha) \ketbra{\alpha}$, and thus cannot show any genuine quantum effects.

The observed differences in the HHG spectrum between the distinct quantum states of light are thus purely due to the fact that the statistical distribution of the intensities are distinct for different states.
In other words, the enhanced HHG yield is simply due to the fact that the ``quantum'' fields have large fluctuations in intensity for the same average intensity.
A classical source of pulses with varying peak intensities would give the same result.
Other light states do not possess an inherent advantage, excluding the possibility of non-classical effects in the obtained yield under the approximations discussed.

\subsection{Many-atom correction}
\begin{figure}[tb]
    \includegraphics[width=\linewidth]{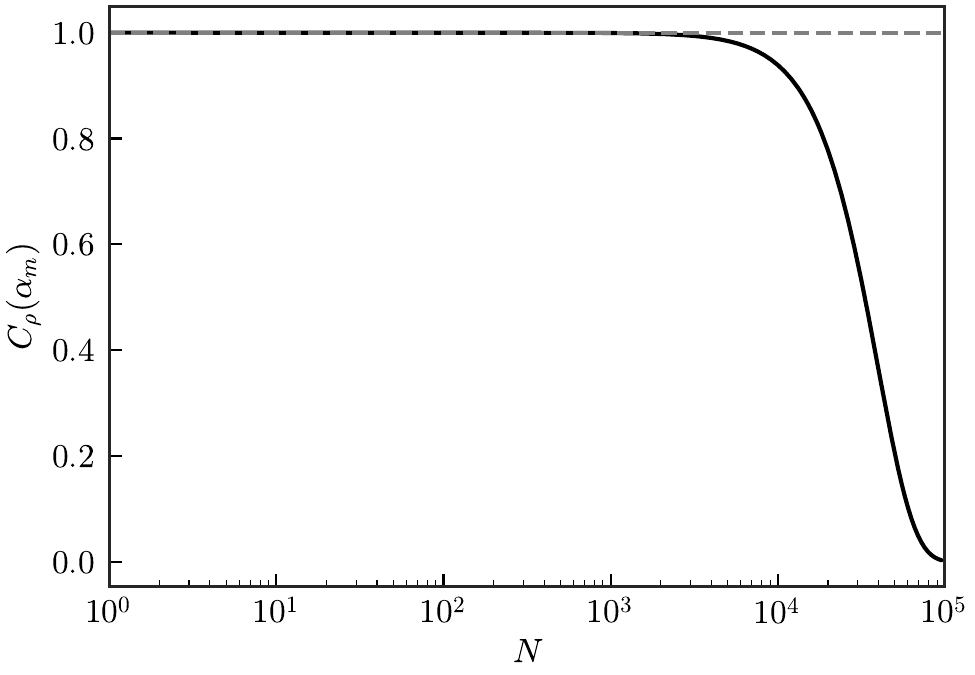}
    \caption{Correction factor $C_\rho(\alpha_m,\delta\alpha)$, given by \autoref{eq:c_rho}, of the trace of the density matrix of the system as a function of the number of atoms $N$. The deviation from unity at large $N$ indicates a breakdown of the mean-field approximation.}
    \label{fig:tr_rho}
\end{figure}
As discussed in the theory section, the mean-field coherent-response approximation can break down, leading to a loss of unitarity, in the many-atom regime due to the exponential scaling of the overlap correction factor $f_{\text{ov}}(\alpha_m,\delta\alpha)$ with the number of atoms $N$ and its square.
However, this does not imply a breakdown of the diagonal approximation, which is only affected by the dipole correction factor $\vec{d}_{\alpha}(\omega) \cdot \vec{d}^*_{\beta}(\omega) / |\vec{d}_{\alpha_m}\!(\omega)|^2$, which is independent of $N$.
We thus evaluate $C_\rho(\alpha_m)$ as given in \autoref{eq:c_rho} to quantify the loss of unitarity in the many-atom regime. In order to do so for arbitrary values of $N$, we express the overlap correction factor, \autoref{eq:fov_many_atom}, as $f_{\text{ov}}(\vec{\alpha}_m,\delta\vec{\alpha}) = e^{F(\alpha_m,\delta\alpha)}$, and perform a Taylor expansion of $F(\alpha_m,\delta\alpha)$ around $\delta\alpha = 0$ up to second order (with the required Jacobian and Hessian matrices evaluated numerically), which is justified by the Gaussian suppression of the integral in \autoref{eq:c_rho}. The trace correction factor $C_\rho$ is then given by a standard Gaussian integral that can be solved analytically, and the scaling of the involved terms with $N$ and $N^2$ can be trivially accounted for without requiring additional numerical calculations.

The resulting trace correction factor $C_\rho(\alpha_m)$ is shown in \autoref{fig:tr_rho} as a function of $N$. As expected, for small $N$ the correction factor is very close to unity, indicating that the mean-field coherent-response approximation is valid in this regime. However, as $N$ increases, the correction factor deviates significantly from unity, indicating a breakdown of the mean-field approximation and a loss of unitarity in the dynamics. 
More detailed inspection of the different terms contributing to the exponent $F(\alpha_m,\delta\alpha)$ reveals that the observed suppression for $N \gtrapprox 10^4$ is primarily due to the atomic overlap $\braket*{\phi_{\beta}}{\phi_{\alpha}}$, which in turn is dominated by phase differences in the ground state components of the atomic wavefunctions at the end of the pulse. These phase differences are caused by the AC Stark shift of the ground state energy, which during the propagation accumulate to give slightly different final phases for different driving field strengths, i.e., for $\delta\alpha \not= 0$.
This observation shows that the mechanism for the breakdown of the mean-field approximation shown here is different from the one that is commonly discussed in the context of strong-field physics (when the incoherent part of emission becomes important due to dipole transitions between different atomic states~\cite{lewenstein1994,gorlach2020}).
We mention that in a recent pre-print, Gothelf et al.~\cite{gothelf2026} have shown that the diagonal approximation induces errors for describing relative measures (quadrature squeezing and photon statistics) in HHG for a one-band model.
These errors increase with the density of emitters, which is in agreement to our results and demonstrate that the correction factors derived here are relevant beyond the average photon-number observable.

\section{Conclusions}
\label{sec.Summary}

In this work, we have extended the theoretical framework for high-harmonic generation (HHG) driven by quantum states of light from the single plane-wave mode treatment of Ref.~\cite{gorlach2023} to a completely general multi-temporal-mode description. This extension enables the exploration of quantum-statistics effects under driving by any multimode quantum state of light, including arbitrary quantum correlations between the modes. At the same time, we have derived a correction factor to the HHG spectrum (easily extensible to any observable of interest) that quantifies the validity of the diagonal approximation, which is a key assumption in the previous treatment. This correction factor is independent of the quantum state of light (represented by its Husimi distribution in the current formalism). Finally, this formalism resolves conceptual inconsistencies in the previous treatment by removing the dependence on an infinitely extended, non-normalizable plane wave and bringing the analytical framework into full agreement with numerical simulations employing finite pulses.

We have shown that the temporal-mode expansion introduces a correction factor $C_\omega(\vec{\alpha}_m)$ to the HHG spectrum. For a single temporal-mode and peak intensities of $I_\mathrm{HHG} \sim 10^{14}$~W/cm$^2$ typical of HHG experiments, this correction factor deviates from unity by less than $10^{-4}$ across the entire HHG spectrum, validating the diagonal approximation and confirming the accuracy of the single-mode result for single-atom HHG driven by pulses in arbitrary quantum states in free space.

Our results demonstrate that the HHG spectrum can be computed by averaging semi-classical calculations over the Husimi quasi-probability distribution $Q(\alpha)$ of the driving field. The validity of the diagonal approximation also implies that convolving this distribution with a Gaussian of unit width in $\alpha$ does not significantly change the HHG spectrum. Since such a convolution corresponds to a classical statistical mixture of coherent states, genuine quantum effects cannot be observed in HHG under these conditions. The enhanced yield observed for states such as bright squeezed vacuum arises purely from their broader statistical distribution of peak intensities, not from quantum effects. A classical ensemble of pulses with the same intensity distribution would produce identical results.

Moreover, we have found that the effect of field fluctuations on the collective ground state of large ensembles of atoms ($N\gg10^4$) leads to nonphysical effects, in particular a breakdown of unitarity and a suppression of the trace of the system's density matrix. This is a direct consequence of the mean-field approximation and of assuming independent emitters. By neglecting these fluctuations, the diagonal approximation recovers the unitarity of the dynamics. As such, our results demonstrate the importance of including the quantum correlations of the dipoles and the role of the incident field in mediating atomic interactions for an accurate fully-quantum account of collective emission.

Finally, the fundamental reason for the absence of quantum effects in single-atom free-space HHG is the requirement of large photon numbers ($|\alpha_m|^2 \sim 10^{11}$). Under such conditions, quantum fluctuations $\delta\alpha \sim 1$ are negligible compared to $\alpha_m \sim 10^5$, and the diagonal approximation becomes exact. However, in nanophotonic environments with ultrasmall mode volumes ($V \sim 1$~nm$^3$), reaching HHG intensities would require only $\sim 10$ photons. In such few-photon regimes, the diagonal approximation breaks down, and genuine quantum effects could potentially manifest, opening new directions for exploring quantum strong-field physics at the intersection of quantum optics and attosecond science.

\begin{acknowledgments}
This work was funded by the Spanish Ministry of Science, Innovation and Universities-Agencia Estatal de Investigación through Grants PID2021-125894NB-I00, CNS2023-145254, PID2024-161142NB-I00 and CEX2023-001316-M (through the María de Maeztu program for Units of Excellence in R\&D). We also acknowledge financial support from the European Union's Horizon Europe Research and Innovation Programme through grant agreement 101070700 (MIRAQLS).
\end{acknowledgments}

\appendix
\section{Derivation of Schrödinger Equation}\label{app:schrodinger}
The general solution of the von Neumann equation, \autoref{eq:rho_alphabeta(t)}, may be formally written with the time-evolution operator $U(t)$
\begin{equation}
    \begin{split}
        \rho_{\alpha \beta}(t) &= U(t)\rho_{\alpha \beta}(0)U^{\dagger}(t), \\
        U(t) &= \mathcal{T} e^{-\frac{i}{\hbar}\int^{t}_0 \dd{\tau}\hat{H_o}(\tau)},
    \end{split}
    \label{eq:Appendix1}
\end{equation}
with $\mathcal{T}$ being the time-ordering operator. It is now convenient to introduce the displacement operator $D(\vec{\alpha})$. This operator displaces a state by the complex amplitude $\vec{\alpha}$ in phase space, and thus creates a coherent state when applied to vacuum, $D(\vec{\alpha})\ket{0} = \ket{\vec{\alpha}}$. Using displacement operators, we can rewrite $\rho_{\vec{\alpha} \vec{\beta}}(0) $ as $\rho_{\vec{\alpha} \vec{\beta}}(0) = \ket{g}\bra{g} \otimes D(\vec{\alpha}) \prod\limits_{\nu} \ket{0_{\nu}}\bra{0_{\nu}} D^{\dagger}(\vec{\beta})$. Additionally, a new operator $\tilde{\rho}_{\vec{\alpha}\vec{\beta}}(t)$ can be defined as $\tilde{\rho}_{\vec{\alpha}\vec{\beta}}(t) = D^{\dagger}(\vec{\alpha})\rho_{\vec{\alpha}\vec{\beta}}(t)D(\vec{\beta}) $, given explicitly by
\begin{equation}
\begin{aligned}
    \tilde{\rho}_{\vec{\alpha} \vec{\beta}}(t) = & D^{\dagger}(\vec{\alpha})U(t)D(\vec{\alpha}) \ket{g}\bra{g} \otimes \\
    &\otimes \prod\limits_{\nu} \ket{0_{\nu}}\bra{0_{\nu}} D^{\dagger}(\vec{\beta}) U^{\dagger}(t) D(\vec{\beta}).
    \label{eq:rho_tilde}
    \end{aligned}
\end{equation}
This operator can be expressed as $\tilde{\rho}_{\vec{\alpha} \vec{\beta}}(t) = \ketbra{\psi_{\vec{\alpha}}(t)}{\psi_{\vec{\beta}}(t)}$ by defining
\begin{equation}
    \ket{\psi_{\vec{\alpha}}(t)} = D^{\dagger}(\vec{\alpha})U(t)D(\vec{\alpha}) \ket{g}\otimes \prod\limits_{\nu} \ket{0_{\nu}},
    \label{eq:psi_alpha2}
\end{equation}
which allows us to rewrite \autoref{eq:rho(t)} as
\begin{equation}
    \rho(t) = \int{\dd[2]{\vec{\alpha}} \dd[2]{\vec{\beta}} \frac{P(\vec{\alpha}, \vec{\beta}^*)}{\bra{\vec{\beta}}\ket{\vec{\alpha}}}D(\vec{\alpha})\ketbra{\psi_{\vec{\alpha}}(t)}{\psi_{\vec{\beta}}(t)}D^{\dagger}(\vec{\beta}) }.
\end{equation}
The trace of $\rho(t)$ can be straightforwardly checked to be equal to 1, as is necessary for probability to be conserved.

Using Eqs.~\eqref{eq:rho_tilde} and~\eqref{eq:psi_alpha2}, \autoref{eq:rho_alphabeta(t)} can be expressed as
\begin{equation}
\begin{aligned}
    & i\hbar \pdv{}{t} \left( D(\vec{\alpha})\ketbra{\psi_{\vec{\alpha}}(t)}{\psi_{\vec{\beta}}(t)}D^{\dagger}(\vec{\beta}) \right)   = \\
    & = [{H_o}(t), D(\vec{\alpha})\ketbra{\psi_{\vec{\alpha}}(t)}{\psi_{\vec{\beta}}(t)}D^{\dagger}(\vec{\beta})].
    \end{aligned}
\end{equation}
After operating, we find that $\ket{\psi_{\vec{\alpha}}(t)}$ satisfies the Schrödinger equation
\begin{equation}
    i\hbar\pdv{\ket{\psi_{\vec{\alpha}}(t)}}{t} =D^{\dagger}(\vec{\alpha}){H_o}(t)D(\vec{\alpha})\ket{\psi_{\vec{\alpha}}(t)},
    \label{eq:psia_psib}
\end{equation}
and so does $\ket{\psi_{\vec{\beta}}(t)}$ under the replacement $\vec{\alpha}\to\vec{\beta}$.
Inserting $H_o(t) = H_A(t) +  \vec{d} \cdot \vec{E}(x,t)$, we can use the fact that the displacement operators do not act on the atomic Hamiltonian $H_A(t)$, and that they are unitary $D(\vec{\alpha})D^{\dagger}(\vec{\alpha}) = 1$, to show that only the second term is affected. The dipole operator $\Vec{d}$ remains unchanged as it is an atomic operator, but the displacement operators do act on the electric field $\Vec{E}(x,t)$. Using the properties $D^{\dagger}(\vec{\alpha}) a_\nu D(\vec{\alpha}) = a_\nu + \alpha_\nu$, and $D^{\dagger}(\vec{\alpha}) a_\nu^{\dagger} D(\vec{\alpha}) = a_\nu^\dagger + \alpha_\nu^*$, we get that
\begin{equation}
    D^{\dagger}(\vec{\alpha}) \Vec{E}(x,t) D(\vec{\alpha}) = \Vec{E}(x,t) + \Vec{E}_{\vec{\alpha}}(x,t),
\end{equation}
where $\Vec{E}_{\vec{\alpha}}(x,t)$ is the classical electric field corresponding to the coherent state $\ket{\vec{\alpha}}$, given by $\Vec{E}_{\vec{\alpha}}(x,t) = \sum_{\nu} \vec{f}_\nu(x,t) \alpha_\nu + \mathrm{H.c.}$.
This way, we can rewrite \autoref{eq:psia_psib} with a ``new'' Hamiltonian with two electric-field interaction terms, one for the quantized field and one for the classical field:
\begin{equation}
    \begin{split}
        i\hbar\pdv{\ket{\psi_{\vec{\alpha}}(t)}}{t} &= H_{\vec{\alpha}}(t) \ket{\psi_{\vec{\alpha}}(t)}, \\
        H_{\vec{\alpha}}(t) &= {H}_A + {\Vec{d}} \cdot {\Vec{E}} + {\Vec{d}} \cdot {\Vec{E}}_{\vec{\alpha}}(t),
    \end{split}
\end{equation}
with initial condition $\ket{\psi_{\vec{\alpha}}(0)} = \ket{g} \otimes \prod\limits_{\nu}\ket{0_{\nu}}$.

\section{Fourier Transform of the Electric Field}\label{app:E_field_transform}

In order to simplify the second part of the exponential in the correction factor in \autoref{eq:Correction_factor}, $t_2(\vec{\alpha},\vec{\beta}) = - \tilde{\vec{\alpha}}^* \cdot \delta\vec{\gamma} + \tilde{\vec{\beta}} \cdot \delta\vec{\gamma}^*$, we consider its first term and express it in terms of the Fourier transform of the electric field. We start by writing $\tilde{\vec{\alpha}}^* \cdot \delta\vec{\gamma}$ as an explicit sum over modes and inserting the definition of the $\gamma$ coefficients, \autoref{eq:gamma_coherent}, to obtain
\begin{equation}
    \tilde{\vec{\alpha}}^* \cdot \delta\vec{\gamma}
    = \sum_\ell \tilde{\alpha}_\ell^* \delta\gamma_\ell
    = -\frac{1}{\hbar} \sum_\ell \tilde{\alpha}_\ell^* \vec{f}_{\ell}(\vec{0}) \cdot \delta\vec{d}(\omega_\ell).
\end{equation}
We now write the positive-frequency part of the classical electric field at the atomic position in the plane-wave basis as $\Vec{E}^+_{\vec{\alpha}}(t) = \sum_\ell \tilde{\alpha}_\ell \vec{f}_\ell(\vec{0}) e^{-i\omega_\ell t}$. Its Fourier transform (which is only nonzero for $\omega > 0$, as indicated by the name ``positive-frequency'') is given by
\begin{multline}
    \Vec{E}^+_{\vec{\alpha}}(\omega) = \sum_\ell \tilde{\alpha}_\ell \vec{f}_\ell(\vec{0}) \int\limits^{+\infty}_{-\infty}\dd{t}\frac{e^{i (\omega - \omega_\ell) t}}{\sqrt{2\pi}} \\ 
    = \sqrt{2\pi} \sum_\ell \tilde{\alpha}_\ell \vec{f}_\ell(\vec{0}) \delta(\omega - \omega_\ell).
\end{multline}
This allows us to rewrite the sum over modes as an integral over frequencies, giving
\begin{equation}
    \tilde{\vec{\alpha}}^* \cdot \delta\vec{\gamma} = \frac{1}{\hbar\sqrt{2\pi}} \int_0^{\infty} \dd{\omega} \Vec{E}^{+*}_{\vec{\alpha}}(\omega) \cdot \delta\vec{d}(\omega),
\end{equation}
where we used that $\vec{f}_\ell(\vec{0}) = -\vec{f}_\ell^*(\vec{0})$, and inserting the analogous expression for $\tilde{\vec{\beta}} \cdot \delta\vec{\gamma}^*$ then gives \autoref{eq:Second_term_reworked}.

\section{Normalization of field}\label{app:normalization}

We here discuss how to normalize the electric field for a single temporal mode $\mathrm{p}$ with a given temporal shape. From Ref.~\cite{blow1990continuum}, a transversal field that is confined within a cross-sectional area $A$ but extends to infinity along one axis, with a certain polarization, may be expressed as
\begin{equation}
    E_{1,\mathrm{p}}^{+}(x,t) = i \int_0^{\infty} \dd{\omega} \left( \frac{\hbar \omega}{4 \pi \varepsilon_0 c A} \right)^{1/2} c(\omega) e^{ -i \omega \left( t - \frac{x}{c} \right)},
    \label{eq:1_photon_field}
\end{equation}
where we emphasize the subindex 1 to make clear that this is the quantized field of a single photon, and correct normalization requires $\int_0^{\infty} \dd{\omega} |c(\omega)|^2 = 1$.
Since we are only interested in $x = 0$ we can evaluate it at that particular $x$ and have it be a function of time only.
Writing the field as an inverse Fourier transform gives
\begin{equation}
\begin{aligned}
    E_{1,\mathrm{p}}^{+}(0,t) &= i \int_0^{\infty} \dd{\omega} \left( \frac{\hbar \omega}{4 \pi \varepsilon_0 c A} \right)^{1/2} c(\omega) e^{ -i \omega t } = \\ & = \frac{1}{\sqrt{2\pi}} \int_0^{\infty} \dd{\omega} \tilde{E}_{1,\mathrm{p}}^+(\omega) e^{ -i \omega t },
    \end{aligned}
\end{equation}
establishing the relation $c(\omega) = -i \tilde{E}_{1,\mathrm{p}}^+(\omega) \left(\frac{2 \varepsilon_0 c A}{\hbar \omega} \right)^{1/2}$.
The normalization of $c(\omega)$ then requires solution of the integral $\int_0^{\infty} \dd{\omega} \frac{|\tilde{E}^+(\omega)|^2}{\omega}$, which for the flat-top laser pulse considered in this work, given by \autoref{eq:Electric_field}, can be expressed as
$\frac{E_{1,\mathrm{p}}^2}{\omega_0^2} \mathcal{I}(n_r,n_f)$, where $\mathcal{I}(n_r,n_f)$ is a unitless normalization integral whose value depends on the number of laser cycles in the ramp-on ($n_r$) and constant ($n_f$) parts of the pulse. Although it can be solved analytically for integer values of $n_r$ and $n_f$, the expression is quite lengthy and not given here for conciseness. For the values used in this work, $n_r = 5$ and $n_f = 15$, we find $\mathcal{I}(5,15) \approx 28.814$.
We note that the specific pulse shape chosen is only normalizable for integer values of $n_r$ and $n_f$, indicating that it does not correspond to a physical pulse propagating in free space for other values of these parameters. This is because it would break the condition that the value of the vector potential is the same before and after the pulse that is required for propagating pulses.
Using this integral, we obtain the electric field amplitude $E_{1,\mathrm{p}}$ of the quantized temporal mode, as given in \autoref{eq:Electric1_field},
\begin{equation}
    E_{1,\mathrm{p}} = \sqrt{\frac{\hbar \omega_0}{2 \varepsilon_0 V_{\text{eff}}}},
\end{equation}
where $V_{\text{eff}} = n_{\text{eff}} A \lambda_0$ is an effective mode volume, with $n_{\text{eff}} = \frac{2}{\pi} \mathcal{I}(n_r,n_f) \approx 18.343$ being the effective number of cycles in the pulse. Note that $n_{\text{eff}}$ is close to, but not exactly equal to $n_f + \frac23 n_r = 18.333$, the value that would be obtained when replacing $\omega\to\omega_0$ in the denominator of the normalization integral.

\bibliography{references}

\end{document}